\documentclass[letter]{aa} 
\usepackage{graphicx}
\usepackage{txfonts}
\usepackage{wrapfig, subfig, threeparttable, rotating}
\usepackage{natbib, lscape}   
\bibpunct{(}{)}{,}{a}{}{,} 
\begin{document}

   \title{First spectrally-resolved H$_2$ observations towards 
HH~54\thanks{Based on observations made with ESO telescopes at the La Silla Paranal Observatory under programme IDs: 089.C-0772, 292.C-5025} 
   }

   \subtitle{Low H$_2$O abundance in shocks}

   \author{G. Santangelo
          \inst{1,2}
          \and
          S. Antoniucci
          \inst{2}
          \and
          B. Nisini
          \inst{2}
          \and
          C. Codella
          \inst{1}
          \and
          P. Bjerkeli
          \inst{3,4,5}
          \and
          T. Giannini
          \inst{2}
          \and
          A. Lorenzani
          \inst{1}
          \and
          L.~K. Lundin
          \inst{6}
          \and
          S. Cabrit
          \inst{7}
          \and
          L. Calzoletti
          \inst{8}
          \and
          R. Liseau
          \inst{5}
          \and
          D. Neufeld
          \inst{9}
          \and
          M. Tafalla
          \inst{10}
          \and
          E.~F. van Dishoeck
          \inst{11,12}
          }

   \institute{Osservatorio Astrofisico di Arcetri, Largo Enrico Fermi 5, 
              I-50125 Florence, Italy \\
              \email{gina@arcetri.astro.it}
         \and
              Osservatorio Astronomico di Roma, via di Frascati 33, 00040 Monteporzio Catone, Italy
         \and         
              Niels Bohr Institute, University of Copenhagen, Juliane Maries Vej 30, DK-2100 Copenhagen {\O}., Denmark
         \and
              Centre for Star and Planet Formation and Natural History Museum of Denmark, University of Copenhagen, {\O}ster Voldgade 5--7, DK-1350 Copenhagen K., Denmark
         \and
              Department of Earth and Space Sciences, Chalmers University of Technology, Onsala Space Observatory, 439 92 Onsala, Sweden
         \and
              European Southern Observatory, Karl Schwarzschild str.2, D-85748 Garching bei Muenchen, Germany
         \and
              LERMA, Observatoire de Paris, UMR 8112 of the CNRS, 61 Av. de l'Observatoire, 75014 Paris, France
         \and
              ASDC, I-00044 Frascati, Roma, Italy
         \and
              The Johns Hopkins University, Baltimore, MD 21218, USA
         \and
              Observatorio Astron\'omico Nacional (IGN), Alfonso XII 3, E-28014 Madrid, Spain
         \and
              Leiden Observatory, Leiden University, P.O. Box 9513, 2300 RA Leiden, the Netherlands
         \and
              Max Planck Institut für Extraterrestrische Physik (MPE), Giessenbachstr.1, D-85748 Garching, Germany
             }

   \date{Received August 5, 2014; accepted September 7, 2014}

 
  \abstract
   {\emph{Herschel} observations suggest that the H$_2$O 
distribution in outflows from low-mass stars resembles the H$_2$ emission. 
It is still unclear  which of the different excitation components 
that characterise the mid- and near-IR H$_2$ distribution is associated with H$_2$O.
}
   {The aim is to spectrally resolve the different excitation components 
observed in the H$_2$ emission. This will allow us to identify the H$_2$ counterpart 
associated with H$_2$O and finally derive directly an H$_2$O abundance estimate with respect to H$_2$.
}
   {We present new high spectral resolution observations of H$_2$ 0-0 S(4), 
0-0 S(9), and 1-0 S(1) towards HH~54, a bright nearby shock region in the southern sky.
In addition, new \emph{Herschel}-HIFI H$_2$O (2$_{12}$$-$1$_{01}$) 
observations at 1670~GHz are presented. 
}
   {Our observations show for the first time a clear separation in velocity 
of the different H$_2$ lines: the 0-0 S(4) line at the lowest excitation peaks at $-$7~km~s$^{-1}$, 
while the more excited 0-0 S(9) and 1-0 S(1) lines peak at $-$15~km~s$^{-1}$.
H$_2$O and high-$J$ CO appear to be associated with 
the H$_2$ 0-0 S(4) emission, which traces a gas component with a temperature of 700$-$1000~K.
The H$_2$O abundance with respect to H$_2$ 0-0 S(4) is estimated 
to be $X$(H$_2$O)$<$1.4$\times$10$^{-5}$ in the shocked gas over an area of 13$^{\prime\prime}$.
}
   {We resolve two distinct gas components associated with the HH~54 shock region 
at different velocities and excitations. 
This allows us to constrain the temperature of the H$_2$O emitting gas ($\leq$1000~K) and 
to derive correct estimates of H$_2$O abundance in the shocked gas, 
which is lower than what is expected from shock model predictions.
}

   \keywords{Stars: formation -- Infrared: ISM -- ISM: jets and outflows -- ISM: Herbig-Haro objects -- ISM: individual objects: HH~54 
               }

   \maketitle
%

\section{Introduction}

Protostellar jets and outflows are a direct consequence of the accretion mechanism in young stellar objects 
during their earliest phase \citep[e.g.][]{ray2007}.
The interaction between the ejecta and the circumstellar medium occurs via radiative shocks
\citep[e.g.][]{kaufman1996,flower2010}, 
whose energy is radiated away through emission lines of atomic, ionic, and molecular species. 
Hot gas at temperatures above 2000~K cools principally through H$_2$ ro-vibrational lines 
in the near-IR and abundant atomic and ionic species \citep[e.g.][]{eisloeffel2000,giannini2004}.   
Warm gas components at hundreds of Kelvin cool via mid- and far-IR molecular lines, 
particularly rotational transitions of H$_2$ (at $\lambda$$\le$28~$\mu$m) 
and lines of other molecular species, such as CO and H$_2$O.

Water has a key role in protostellar environments \citep{vandishoeck2011}. 
Its abundance with respect to H$_2$ 
is expected to increase from $<$10$^{-7}$ in cold regions 
up to 3$\times$10$^{-4}$ in warm gas due to the combined effects of sputtering of grain mantles 
and high-temperature reactions 
\citep[][]{hollenbach1989,kaufman1996,flower2010,suutarinen2014}.
The \emph{Herschel} Space Observatory revealed the complexity of H$_2$O line profiles 
\citep[e.g.][]{codella2010,kristensen2012,santangelo2012,vasta2012} 
and showed that H$_2$O emission probes warm ($\gtrsim$300~K) and 
dense ($n_{\rm H_2}$$>$10$^5$~cm$^{-3}$) gas with spatial distribution 
that resembles the H$_2$ emission \citep[e.g.][]{nisini2010,tafalla2013,santangelo2013}. 
Low H$_2$O abundances are derived in outflows for warm shocked gas, 
ranging from a few $\times$10$^{-6}$ to a few $\times$10$^{-5}$
\citep[e.g.][]{bjerkeli2012,santangelo2013,tafalla2013,busquet2014}. 
These abundance values 
are at least an order of magnitude lower than what is expected in warm shocked gas from shock model
predictions \citep[e.g.][]{kaufman1996,flower2010}.
Their determinations rely on the assumption that H$_2$O 
traces the same gas as the spectrally unresolved low-$J$ H$_2$ 0-0 lines.
Spectrally resolved observations of H$_2$ 
are thus needed to directly compare the line profiles and finally test this hypothesis.

The Herbig-Haro object HH~54 is located in the nearby star-forming region Chamaeleon~II
\citep[$D$$=$180~pc,][]{whittet1997}. 
The object shows a clumpy appearance, consisting of several arcsecond-scale bright knots.  
\citet{knee1992} associates HH~54 with a monopolar blue-shifted CO outflow, 
whose driving source remains unclear \citep[e.g.][]{caratti2006,ybarra2009,bjerkeli2011}.   
Mid-IR cooling is dominated by pure rotational H$_2$ lines   
\citep{cabrit1999,giannini2006,neufeld1998,neufeld2006}
probing warm gas with a mixture of temperatures in the range 400$-$1200~K.
HH~54 was also observed in several lines of CO and H$_2$O from space 
and the ground \citep{liseau1996,nisini1996,bjerkeli2009,bjerkeli2011}.

In this letter we present new ESO VLT high-resolution spectroscopic observations of H$_2$ 
towards HH~54.
The observations are complemented with 
\emph{Herschel}-HIFI observations of H$_2$O (2$_{12}$$-$1$_{01}$). 
This unique dataset is used to spectrally resolve the different excitation components 
observed in H$_2$. We are finally able to identify the H$_2$ counterpart associated 
with H$_2$O and derive the H$_2$O abundance in the shocked gas directly.


\section{Observations}
\label{sect:observations}

\begin{figure}
\centering
\includegraphics[width=0.35\textwidth]{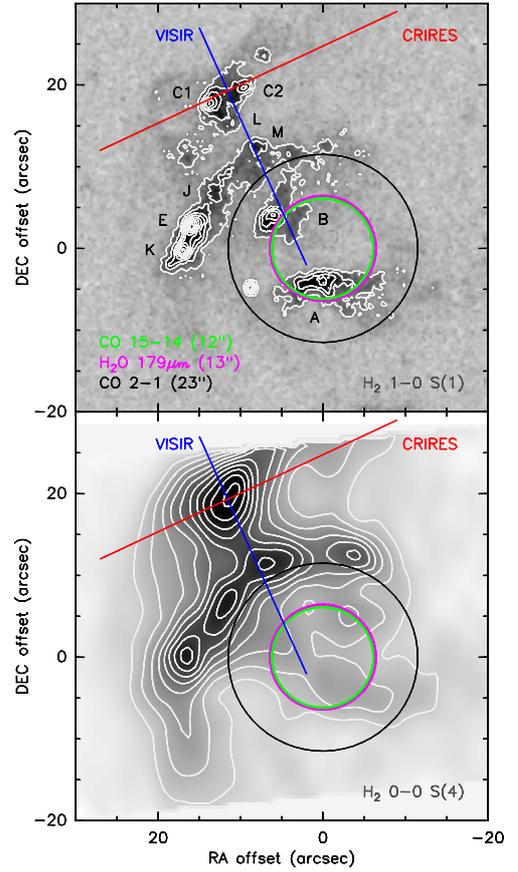}   
\caption{\emph{Upper:} H$_2$ 1-0 S(1) image of HH~54 
from NTT-SofI observations \citep{giannini2006}. 
The positioning of the slits adopted for VLT-VISIR and CRIRES observations 
is shown in blue and red. The beam sizes of \emph{Herschel} 
H$_2$O (2$_{12}$$-$1$_{01}$) (magenta circle), CO (15$-$14) (green, Bjerkeli et al. submitted), and SEST CO (2$-$1) 
\citep[black,][]{bjerkeli2009,bjerkeli2011} are displayed. 
Offsets are relative to: $\alpha_{J2000}$$=$12$^{\rm h}$55$^{\rm m}$50$.\!\!^{s}$3, 
$\delta_{J2000}$$=$$-$76$^\circ$56$^{\prime}$23$^{\prime\prime}$ \citep{bjerkeli2011}.
\emph{Lower:} Spitzer-IRS H$_2$ 0-0 S(4) image of HH~54 \citep{neufeld2006}.  
Symbols are the same as in the upper panel.
}
\label{fig:puntamenti}
\end{figure}

Our dataset consists of data collected towards HH~54 
with ESO facilities (Table~\ref{table:observations}) and with \emph{Herschel}.
Figure~\ref{fig:puntamenti} 
shows the VLT slit positions and \emph{Herschel} and SEST beam sizes for
the observations presented in this paper 
in comparison with the H$_2$ 1-0 S(1) and 0-0 S(4) maps of the region 
\citep{giannini2006,neufeld2006}.

\subsection{VISIR high-resolution mid-IR spectroscopy}
\label{subsect:VISIR}

\begin{table*}
\caption{H$_2$ transitions observed.}
\label{table:observations}
\centering
\begin{tabular}{l c c c c c c c c r }
\hline\hline
Transition & Wavelength & $E_{\rm up}$ & Telescope/Instrument & Slit & PA & $R$=$\lambda/\Delta\lambda$ & $\Delta$$v$ & Date & Exposure time \\
& ($\mu$m) & (K) & & ($^{\prime\prime}$) & ($^{\circ}$) & & (km~s$^{-1}$) & (m, y) & (min) \\
\hline
H$_2$ 0-0 S(4) & 8.0251 &  3474 & VLT/VISIR  &  0.4$\times $32 & $-$26 & 32000 & 10 &     Apr 2012 &  45  \\ 
H$_2$ 0-0 S(9) & 4.6947 & 10263 & VLT/CRIRES &  0.4$\times $40 &   113 & 50000 &  6 & Jan/Feb 2014 & 102  \\
H$_2$ 1-0 S(1) & 2.1218 &  6956 & VLT/CRIRES &  0.4$\times $40 &   113 & 50000 &  6 & Jan/Feb 2014 &  10  \\
\hline
\end{tabular}
\end{table*}

On April 2012 we performed spectrally-resolved observations of H$_2$ 0-0 S(4) 
(see Table~\ref{table:observations}) with VLT-VISIR \citep{lagage2004}.
The 0$.\!\!^{\prime\prime}$4$\times $32$^{\prime\prime}$ slit 
was positioned on the basis of the Spitzer image (see Fig.~\ref{fig:puntamenti}); 
it was oriented in a way to encompass knot B, 
which was covered by the \emph{Herschel} single-pointing observations of H$_2$O 
(see Sect.~\ref{subsect:HIFI}), and the C1/C2 knots, 
which correspond to the brightest knot in the Spitzer H$_2$ emission.
We conducted our observations by chopping and nodding the telescope off-source, 
with equal time on both positions.
Data reduction and calibration were performed by using the VISIR pipeline recipes 
(version 3.5.1)\footnote{https://www.eso.org/sci/software/pipelines/visir/visir-pipe-recipes.html}, 
which provide standard procedures for flat-fielding and background subtraction.
A model for the sky emission lines is used by the pipeline for the wavelength calibration.
To fit the dispersion relation we employed a second degree polynomial, which provides higher 
correlation coefficient with respect to the default pipeline linear solution.
The uncertainty on the peak velocity is about 3~km~s$^{-1}$, comparable with the spectral pixel.
The IRAF package was used for spectra extraction.
Only C1/C2 knots are clearly detected with VISIR; 
knot M is only tentatively detected in the spectral image, 
whereas knot B is not detected.

\subsection{CRIRES high-resolution near-IR spectroscopy}
\label{subsect:CRIRES}

We carried out high-dispersion spectroscopic observations 
of the H$_2$ 0-0 S(9) and H$_2$ 1-0 S(1) transitions 
(Table~\ref{table:observations}) towards HH~54 
with VLT-CRIRES \citep{kaeufl2004}.
Observations were performed between January and February 2014 
during director discretionary time. 
Since only the bright C1/C2 knots were detected by VISIR, 
the CRIRES 0$.\!\!^{\prime\prime}$4$\times $40$^{\prime\prime}$ slit was oriented 
in order to cover them (Fig.~\ref{fig:puntamenti}).
Chopping and nodding were performed along the slit to minimise the integration time.
Data reduction and wavelength calibration were performed with the CRIRES pipeline recipes (version 2.3.1).
The wavelength calibration, based on the comparison with a sky emission model, 
was satisfactory (high correlation coefficient) for the 0-0 S(9).  
OH emission lines were used to refine the wavelength scale for the 1-0 S(1).
The uncertainty associated with peak velocities is $\sim$2.5~km~s$^{-1}$.
The IRAF package was used for spectra extraction.

\subsection{\emph{Herschel}-HIFI observations}
\label{subsect:HIFI}

Single-pointing observations of H$_2$O (2$_{12}$$-$1$_{01}$) at 1669.9~GHz 
were performed with the Heterodyne Instrument for the Far Infrared \citep[HIFI,][]{degraauw2010} 
on board \emph{Herschel} towards HH~54B (see Fig.~\ref{fig:puntamenti}).   
The reference coordinates are $\alpha_{J2000}$$=$12$^{\rm h}$55$^{\rm m}$50$.\!\!^{s}$3, 
$\delta_{J2000}$$=$-76$^\circ$56$^{\prime}$23$^{\prime\prime}$. 
The observations were carried out in September 2012\footnote{The data are part of the OT2 program 
\emph{``Herschel observations of the shocked gas in HH~54''} (observation ID: 1342251604).}.  
The diffraction-limited beam size is $\sim$13$^{\prime\prime}$. 
The data were processed with the ESA-supported package {\sc hipe} version 12.0 for calibration.
The HebCorrection and fitHifiFringe tasks within {\sc hipe} were successfully used to remove the 
electronic standing waves in Band 6, which affected the line.
Further data reduction and analysis 
were performed using the GILDAS\footnote{http://www.iram.fr/IRAMFR/GILDAS/} software.
The antenna temperature scale, $T^*_A$, was converted into the main-beam temperature scale,
$T_{\rm mb}$, using main-beam efficiency factor of 0.71 \citep{roelfsema2012}.
The flux calibration uncertainty is around 10\%, based on cross-calibration with \emph{Herschel}-PACS 
\citep{bjerkeli2011}.
At the velocity resolution of 1~km~s$^{-1}$, the rms noise is 80~mK ($T_{\rm mb}$ scale).


\section{Two velocity components in H$_2$ observations}
\label{subsect:H2}

\begin{figure}
\center
\includegraphics[width=0.43\textwidth]{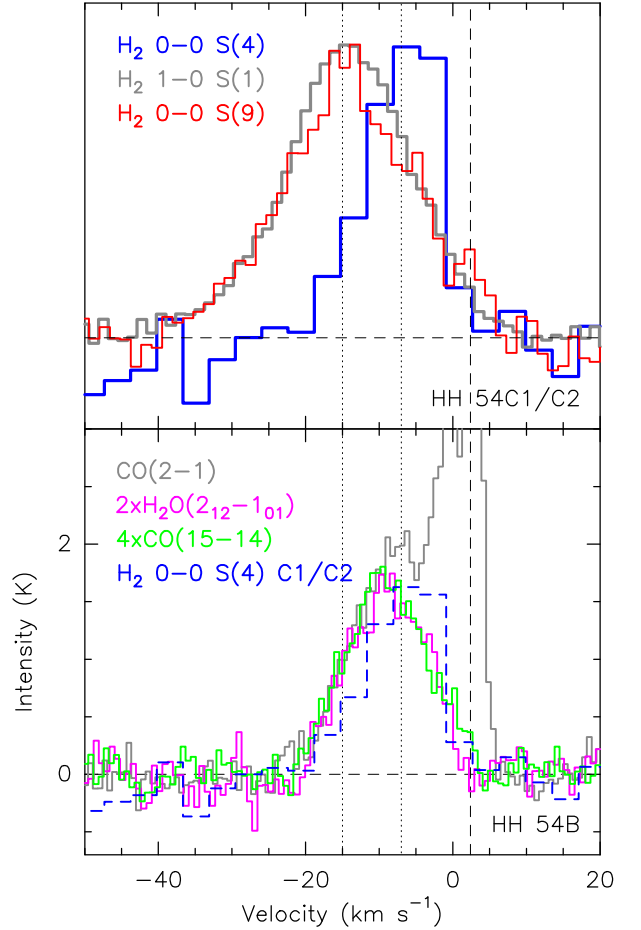} 
\caption{\emph{Upper:} H$_2$ 0-0 S(4) towards HH~54C1/C2 
is compared with 1-0 S(1) and 0-0 S(9). 
Spectra are normalized to their peak values.
The vertical dashed line marks the systemic velocity 
\citep[$v_{\rm LSR}$$=$$+$2.4~km~s$^{-1}$,][]{bjerkeli2011}. 
The two vertical dotted lines indicate the velocity of: the 0-0 S(4) peak at 
$-7$~km~s$^{-1}$; and the 1-0 S(1) and 0-0 S(9) peaks at $-$15~km~s$^{-1}$.
\emph{Lower:} H$_2$O (2$_{12}$$-$1$_{01}$), CO (15$-$14), 
and CO (2$-$1) towards HH~54B are compared with 
H$_2$ 0-0 S(4) at HH~54C1/C2. H$_2$O, CO, and H$_2$ are normalized 
to the peak of the bump feature in CO (2$-$1).
}
\label{fig:spettri}
\end{figure}

Velocity centroids of the CRIRES spectra at C1 and C2 knots are consistent 
within one spectral pixel ($<$3 km/s). The two spectra have thus been averaged 
to compare with VISIR H$_2$ 0-0 S(4) extracted at knot C1/C2. 
The comparison is presented in the upper panel of Fig.~\ref{fig:spettri}.
A peak velocity of $-7$~km~s$^{-1}$ is associated with the 0-0 S(4) line, 
whereas the higher excitation 1-0 S(1) and 0-0 S(9) lines 
peak at the higher blue-shifted velocity of $-15$~km~s$^{-1}$.
Our spectrally-resolved H$_2$ observations clearly show for the first time that 
mid-IR and near-IR H$_2$ 
lines are well separated in velocity, thus representing two distinct velocity components.
This suggests that two separate shock components with different excitation conditions
are associated with gas peaking at different velocities.

The comparison between H$_2$ 0-0 S(4) at HH~54C1/C2 and 
H$_2$O (2$_{12}$$-$1$_{01}$), CO (15$-$14) (Bjerkeli et al. submitted), 
and CO (2$-$1) \citep{bjerkeli2009,bjerkeli2011} observations at HH~54B
is presented in the lower panel of Fig.~\ref{fig:spettri}. 
The low-$J$ CO lines, in particular
CO (2$-$1), present a two-components profile: a triangular-shaped
low-velocity (LV CO, hereafter) component, which peaks 
at the systemic velocity of the cloud ($+$2.4~km~s$^{-1}$);
and an additional superposed ``bump-like'' component \citep{bjerkeli2011} 
centred at the blue-shifted velocity of $-7$~km~s$^{-1}$. 
This latter feature seems to dominate the emission of the high-$J$ CO (15$-$14).
The similarity between CO (15$-$14) and H$_2$O line profiles, 
taken with similar beam sizes (12$^{\prime\prime}$ and 13$^{\prime\prime}$), 
suggests that the bump feature is associated with the H$_2$O emitting gas
and has higher excitation with respect to the LV gas.

Although the H$_2$ 0-0 S(4) spectrum is taken at the C1/C2 knot, 
the comparison with H$_2$O and high-$J$ CO observed at knot B
shows that the three lines trace emission in the same velocity range.
Moreover, taking the different spectral resolutions and the 
uncertainty on the H$_2$ peak velocity determination into account (see Sect.~\ref{subsect:VISIR}), 
the H$_2$ 0-0 S(4) line profile well resembles the H$_2$O and high-$J$ CO line profiles
(Fig.~\ref{fig:spettri}-bottom).
HIFI maps of CO (10$-$9) and lower-$J$ CO lines by \citet{bjerkeli2011} show that, 
although the relative intensity of the LV and bump components changes 
within the HH~54 region, their peak velocities remain constant within 2~km~s$^{-1}$ among 
the different knots. 
We thus also assume that the peak velocity of the H$_2$O emission, 
which appears to be associated with the high-$J$ CO emission at knot B,  
does not change within the region and in particular along the VISIR slit.
In this case, the H$_2$ 0-0 S(4) emission would be 
associated with the same gas as traced by H$_2$O and high-$J$ CO. 

In conclusion, our observations detect for the first time the presence of a stratification in velocity 
in the H$_2$ gas from low- to high-excitation emission lines.
The H$_2$ 0-0 S(4) component appears to be associated with 
H$_2$O and high-$J$ CO, as expected from the comparison between the spatial distributions
\citep[e.g.][]{nisini2010,tafalla2013,santangelo2013}. 
We note that in the low-$J$ CO an additional gas component around the systemic velocity 
is detected. This gas component is not observed in the high-$J$ CO 
lines and in the H$_2$ lines, since higher temperatures are needed to excite them.
On the other hand, the higher velocity component associated with H$_2$ 1-0 S(1) and 0-0 S(9)
is not detected in the CO emission, even in the higher-$J$ lines, 
since it is associated with a gas at even higher temperatures ($T$$\gtrsim$2000~K).


\section{H$_2$O abundance estimate}
\label{subsect:abundance}

Our new observations allow us to spectrally identify the 0-0 S(4) line as 
the H$_2$ counterpart associated with H$_2$O, with the assumption that 
the H$_2$ and H$_2$O profiles do not change between the C and B knots.
This can be used to accurately constrain the temperature of the gas from the 
H$_2$ emission and derive correct H$_2$O abundances with respect to H$_2$.
\citet{neufeld2006} mapped H$_2$ S(0)$-$S(7) pure rotational lines 
towards HH~54 with Spitzer-IRS. 
Their H$_2$ rotational diagram, constructed over a 15$^{\prime\prime}$ region
encompassing HH~54B, indicates the presence of warm gas 
with temperatures in the range 400$-$1200~K. 
According to these authors, a temperature range 
of 700$-$1000~K is associated with the 0-0 S(4) emission, which corresponds 
$N$(H$_2$)$=$6.6$\times$10$^{19}$ and 
2.1$\times$10$^{19}$~cm$^{-2}$ over 13$^{\prime\prime}$ for 700 and 1000~K, respectively.

We assumed for the H$_2$O emission the same temperature range as derived 
from the H$_2$ 0-0 S(4) line and a gas density $n$(H$_2$)$\gtrsim$10$^{5}$~cm$^{-3}$ 
\citep[e.g.][]{tafalla2013,santangelo2013,busquet2014}.
We used the {\sc radex} molecular LVG radiative transfer code \citep{vandertak2007} 
to model the observed H$_2$O (2$_{12}$$-$1$_{01}$) emission. 
A typical line width of 10~km~s$^{-1}$ was adopted from a Gaussian fitting to the spectrum
(see Fig.~\ref{fig:spettri}).
The lower limit on the H$_2$ density corresponds to an upper limit on the derived column density.
In particular, we obtain $N$(H$_2$O)$<$3$\times$10$^{14}$~cm$^{-2}$
over a 13$^{\prime\prime}$ area. 
The comparison with the H$_2$ column density obtained from the 0-0 S(4)  
for the same temperature range gives an H$_2$O abundance $X$(H$_2$O)$<$1.4$\times$10$^{-5}$. 
A lower H$_2$ density of 2$\times$10$^4$~cm$^{-3}$ \citep[submitted]{bjerkeli2011} 
would increase the H$_2$O abundance by a factor of 2.
The upper level energy of H$_2$O (2$_{12}$$-$1$_{01}$), which is about 114~K, 
is much smaller than that of the H$_2$ S(4) line (Table~\ref{table:observations}).
Therefore, we cannot exclude that H$_2$O emission is associated with a colder gas component
that is not probed by our H$_2$ observations. 
However, a temperature lower than the assumed 700$-$1000~K would indicate an even lower 
H$_2$O abundance, thus strengthening our result. 
The derived upper limit on the H$_2$O abundance is in agreement with the 
abundance value of 10$^{-5}$ derived by \citet{liseau1996} and \citet{bjerkeli2011} from ISO and \emph{Herschel} observations 
of transitions at similar wavelengths as well as with the upper limit of $<$1.6$\times$10$^{-4}$
obtained by \citet{neufeld2006} based on non-detections of shorter wavelength transitions covered by Spitzer.
Our H$_2$O abundance estimate in HH~54 confirms the values 
recently found by \emph{Herschel} in outflows from Class 0 sources 
\citep[e.g.][]{bjerkeli2012,santangelo2013,tafalla2013,busquet2014}, 
which are based on the assumption that H$_2$O traces the same gas as traced 
by the low-$J$ H$_2$ emission. 

An estimate of the H$_2$O abundance at the C knot can also be derived 
using the PACS map of H$_2$O (2$_{12}$$-$1$_{01}$) by \citet{bjerkeli2011}.
The H$_2$O flux density at knot C is a factor of 2 lower than at the position of the HIFI H$_2$O 
observations, which yields $N$(H$_2$O)$<$10$^{14}$~cm$^{-2}$. 
The H$_2$ column density obtained from the 0-0 S(4) at knot C 
is in the range $N$(H$_2$)$=$5$\times$10$^{19}$$-$1.6$\times$10$^{20}$~cm$^{-2}$ for 1000 and 700~K, respectively.
The comparison between H$_2$O and H$_2$ indicates an H$_2$O abundance 
$X$(H$_2$O)$<$2$\times$10$^{-6}$, which is even more strict than that derived at knot B.
This indicates a variation of H$_2$O abundance within the HH~54 region, with a decrease towards 
the peak of the H$_2$ S(4) emission. 
This may explain the different emission peaks of the H$_2$O distribution 
observed by PACS \citep{bjerkeli2011} and the H$_2$ S(4) emission.

%

\section{Conclusions}
\label{sect:conclusions}

We present new spectrally-resolved observations towards HH~54 of H$_2$ 0-0 S(4), 0-0 S(9), and 1-0 S(1).
These are complemented by new \emph{Herschel}-HIFI H$_2$O (2$_{12}$$-$1$_{01}$) observations. 
Our data show for the first time the separation in velocity between the gas component traced by 
the low-excitation H$_2$ 0-0 S(4) line and that associated with the H$_2$ lines at higher excitation.
The observed H$_2$ stratification in velocity suggests that our observations resolve 
two distinct gas components associated with the HH~54 shock region at different velocity and excitation.
We spectrally identify the H$_2$ 0-0 S(4) line as the H$_2$ counterpart of H$_2$O emission. 
This allows us to constrain the temperature of the H$_2$O emitting gas ($\leq$1000~K).
H$_2$O abundance is estimated to be lower than what is expected from shock model predictions by 
at least one order of magnitude.
High spectral resolution observations of different targets are needed to confirm this result.

\begin{acknowledgements}
We thank VLT astronomers and operators for performing excellent service mode observations 
at CRIRES and providing excellent support with VISIR.
We particularly thank the ESO Director's Office for the DDT observations with CRIRES.
This work was partly supported by ASI--INAF project 01/005/11/0, PRIN INAF 2012 -- JEDI, 
and Italian Ministero dell'Istruzione, Universit\`a e Ricerca through the grant Progetti Premiali 2012 -- iALMA.
\end{acknowledgements}


\end{document}